# Development of the Complex Nexus of Socio-Techno-Economic-Environmental Parametric (STEEP) Metrics for Evaluating Coal-to-Clean Energy Transitions

Muhammad R. Abdussami[1,*], Aditi Verma[1]

[1]University of Michigan, 2355 Bonisteel Blvd, Ann Arbor, MI 48109, Unites States

**Abstract**:

Transitioning from coal to clean energy (e.g., nuclear and renewables) is essential to mitigate climate change, improve air quality, and ensure sustainable energy security. Reducing reliance on coal lowers greenhouse gas emissions and pollution, enhancing public health and economic growth through renewable energy investments. Additionally, clean energy promotes energy independence and long-term sustainability. This transition involves developing a Complex Nexus of Socio-Techno-Economic-Environmental Parametric (STEEP) Metrics to systematically evaluate and guide these transitions, facilitate informed decision-making, and optimize resource allocation to achieve environmental and socioeconomic benefits. It provides a structured framework for navigating the complexities of repurposing existing coal infrastructure, addressing the multifaceted challenges in aligning societal, technical, economic, and environmental considerations. Thus, the paper presents a comprehensive methodology to assess the feasibility of converting coal plants into clean energy systems. It classifies the methodological approach into three: optimal site selection using a multi-criteria decision-making framework with societal, technical, economic, and environmental criteria to rank suitable coal plant sites; optimal long-term planning evaluation with performance indicators to compare the techno-economic-environmental benefits of different energy mix strategies, such as Greenfield, Coal-to-Nuclear (C2N), and Coal-to-Integrated Energy System (C2IES); and short-term operational benefits assessment via the Unit Commitment Economic Dispatch (UCED) model that optimizes generator scheduling and minimizes costs across multiple scenarios (Coal plant, Greenfield, C2N, and C2IES). The study develops extensive metrics for evaluating the feasibility of coal to clean transitions. This framework enables researchers and practitioners to effectively analyze the potential of this transition and identify the optimal strategies for implementation.

**Highlights:**

- Development of a multi-criteria framework for optimal site selection in coal to clean energy transitions.
- Development of a methodological framework for evaluating the benefits of coal to clean energy transitions from long-term planning perspective.
- Formulation of a methodological architecture for assessing the advantages of coal to clean energy transitions from short-term operational perspective.

**Nomenclature:**

AHP    Analytic Hierarchy Process
CPP    Coal Power Plant
C2N    Coal to Nuclear
C2IES    Coal to Integrated Energy System
EPRI    Electric Power Research Institute
GIS    Geographic Information System
LOLP    Loss of Load Probability
LWR    Light Water Reactor
NRC    Nuclear Regulatory Commission
TES    Thermal Energy Storage
SMR    Small Modular Reactor

1. **Introduction:**

The transition from coal to clean energy is essential because it aligns with the global imperative to decarbonize power generation while ensuring a reliable electricity supply. Coal to clean energy transitions can be conducted in several ways; it could be Greenfield, Coal-to-Nuclear (C2N), and Coal-to-Integrated Energy Systems (C2IES). Greenfield a genuine greenfield nuclear power plant (NPP) construction project, entirely independent of any existing coal power plant (CPP). It does not include any infrastructure or the component from the existing CPP. There are no decommissioning costs associated with CPP; the costs and schedule pertain solely to the NPP construction. C2N option includes different component (e.g., steam-cycle component, electrical component, and transmission lines) and civil infrastructure form the existing CPP. In C2IES, both nuclear reactors and renewable energy sources are deployed into the existing CPP. Components from existing CPP is also reused in C2IES. Coal-to-Renewables option is not considered here since stand-alone renewable system poses significant uncertainty into the system. Replacing coal-fired plants with clean energy leverages existing infrastructure, reduces carbon emissions, and maintains grid stability. However, this shift is inherently complex due to the diverse factors involved. Each potential site must be carefully assessed for compatibility with advanced nuclear reactors and renewable technologies, including cooling water availability, transmission grid integration, seismic safety, and local policy restrictions. Additionally, social acceptance, economic viability, and technical feasibility must be harmonized to ensure the transition minimizes job losses while maximizing community benefits. The process also requires navigating complex regulatory requirements, significant capital investments, and long-term planning to ensure operational success. Thus, while coal to clean conversion is crucial for a sustainable energy future, it presents a multifaceted challenge that demands careful strategic planning.

Selecting optimal sites, conducting long-term planning assessments, and evaluating the short-term operational benefits involve complex computational tasks. The literature on optimal site selection for coal to clean transitions is relatively inadequate. Susiati et al. identify potential nuclear plant sites in West Kalimantan, Indonesia, using GIS and analytical hierarchical processes (AHP). They prioritize coastal locations to support water reactor cooling. The analysis discovers 12 potential sites and recommends two priority locations for their minimal environmental impact and compliance with regional planning guidelines [1]. Devanand at el. propose a novel Mixed Integer Non-Linear Programming (MINLP) model for evaluating land sites suitable for modular nuclear power plants (NPPs). The model considers a range of factors such as cost, cooling water availability, and seismic risk. A case study uses the model to identify optimal modular NPP sites, demonstrating its ability to inform preliminary site analysis [2]. Martins et al. present a comprehensive overview of the criteria and methodologies for selecting suitable sites for nuclear power plants. Emphasizing the importance of integrating GIS and multicriteria decision analysis, it seeks to create transparent and participatory site selection processes that balance safety, technical, economic, and environmental factors [3]. Omitaomu et al. introduce a GIS-based tool, Oak Ridge Siting Analysis for Power Generation Expansion (OR-SAGE). This tool helps energy stakeholders identify optimal locations for advanced nuclear reactors using criteria like population density, water availability, and geological hazards [4]. However, none of the research discusses the criteria for siting while adopting coal to clean energy transitions.

A list of research talks about the economic benefits of C2N transitions. Hansen et al. demonstrated that about 80% of coal plant sites in the U.S. could be retrofitted for advanced nuclear reactors, reducing capital costs by 15-35% and benefiting local economies [5]. Simonian et al. analyzed repurposing Texas's Limestone CPP with advanced reactors, finding the HTGR (Xe-100) most suitable due to its capacity and heat removal abilities [6]. Łukowicz et al. studied replacing a 460 MW Polish coal plant with an SMR, identifying modifications in steam parameters for successful conversion [7]. Bartela et al. compared this Polish plant's retrofit with the Kairos Power Fluoride-salt-cooled High-temperature Reactor (KP-FHR), emphasizing economic benefits despite cost uncertainty [8]. In another study, Bartela et al. showed that incorporating a thermal energy storage (TES) system while repurposing coal plants to nuclear enhances flexibility and profitability [9].

An Idaho National Laboratory (INL) report outlines the technical and economic benefits of converting U.S. coal plants to nuclear, stressing detailed site-specific studies for success [10]. Xu et al. demonstrate a three-stage retrofit strategy with high-temperature nuclear heat sources in China, finding significant cost savings [11]. The Oak Ridge National Laboratory's OR-SAGE tool demonstrated that several Tennessee Valley Authority sites are viable for advanced reactor retrofits due to existing infrastructure and population distribution [12]. However, current research does not provide a strategic methodology for evaluating the long-term planning advantages and short-term operational benefits of retrofitted nuclear plants.

Therefore, this paper includes developing a detailed methodology for Optimal Site Selection, evaluating long-term advantages while repurposing coal plants, and assessing operational benefits when the retrofitted nuclear plants come into operation. The paper is divided into three parts. Firstly, a detailed ranking process employing a multi-criteria decision-making framework is demonstrated to identify the most suitable coal sites for conversion. Next, a planning approach has been developed to evaluate the feasibility and benefits of converting existing coal sites into nuclear or integrated energy systems using key performance indicators and optimization techniques. Lastly, a framework to compare the operational benefits of various energy systems, including coal plants, nuclear greenfield, C2N, and C2IES, has been illustrated. Through these methodologies, the paper aims to provide researchers and practitioners with a robust framework for evaluating and implementing coal to clean energy transitions, ultimately contributing to the global shift toward a cleaner energy future.

2. **Methodology for optimal site selection**

Converting coal plants into nuclear plants requires a detailed feasibility study. Ranking the sites based on feasibility involves analyzing multiple selection criteria. For this type of study, the Analytic Hierarchy Process (AHP) can be a useful method to determine the most feasible coal site for reactor deployment. The AHP is a structured decision-making methodology that helps prioritize criteria by assigning relative importance through pairwise comparisons. The method is beneficial when dealing with complex decisions involving multiple criteria. The steps for implementing AHP into coal to clean energy analysis are illustrated in Fig. 1.

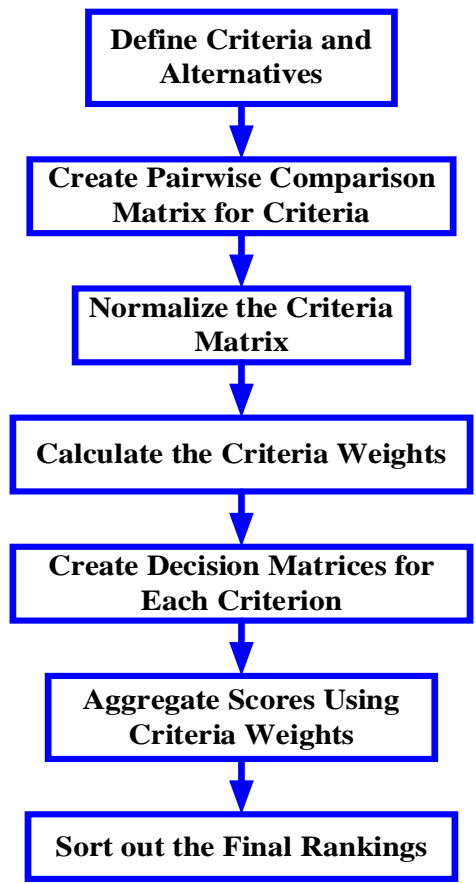

Fig. 1: Steps for Analytic Hierarchy Process (AHP)

The hierarchical model consists of goals, criteria, and alternatives. This study's objective is to rank the identified coal sites for nuclear conversion. The criteria include relevant factors such as water availability, transmission line availability, moratorium restrictions, energy price, and community support. Several societal, technological, economic, and environmental criteria have been developed in this study, which are explained in the later sub-sections. The alternatives are the individual coal sites being considered. To formulate the pairwise matrix, each criterion is compared against every other criterion in pairs, using a numerical scale reflecting relative importance. For example, water availability may be deemed more important than energy price and thus receive a higher score in the comparison matrix. Then, the pairwise comparison matrix is normalized by summing each column and dividing each cell by its column sum. Criteria weights are determined by averaging the normalized scores across each row, providing a numerical value representing each criterion's relative importance. A consistency ratio is computed to ensure the pairwise comparisons are logically consistent. The comparisons are considered reliable if this ratio is below an acceptable threshold (0.10). Each coal site is scored based on its performance in each criterion. Scores can be gathered from objective data or expert judgment and normalized if needed. The final score for each coal site is calculated by multiplying the normalized scores with the corresponding criteria weights and summing across all criteria. It provides a comprehensive score that reflects each site's feasibility for conversion. Coal sites are ranked based on their final scores, highlighting the most suitable candidates for conversion to nuclear power.

For this analysis, societal, technological, economic, and environmental criteria have been outlined, each explained below in terms of its relevance to this study. To incorporate these criteria and associated metrics into an AHP framework, it is necessary to assign numerical values reflecting their relative importance. This assignment of values relies on the expertise of energy system modelers and policymakers. Additionally, specific information (e.g., distance of the substation from the coal plant, regional energy prices associated with the particular coal plant, etc.) is required for each metric corresponding to each coal plant but is not detailed here.

### 2.1. Societal Metrics

| Criteria | Relevance |
|---|---|
| Nuclear restriction | State restrictions on new nuclear reactors, ranging from minor to complete bans, can significantly delay or impede project siting due to required approvals and moratoriums. If a state has restrictions on nuclear siting, the site will be less appropriate for new reactor deployment. |
| Nuclear Inclusive Policy | Incentivizing nuclear energy makes it more cost-competitive in the energy market. State policies like renewable portfolio standards and clean energy goals can play a crucial role in supporting the development of C2N and C2IES projects [13], [14]. |
| Social Vulnerability Index (Buffer distance from site: 10 miles) | The Social Vulnerability Index (SVI) measures how well counties can withstand natural disasters, human-caused disasters, and disease outbreaks [15]. |
| Protected Lands | Protected lands (e.g., critical habitats, hospitals, national monuments, national and local parks, schools/colleges, wild and scenic rivers, forests, wilderness areas, wildlife refuges, and correctional facilities) are often excluded from nuclear siting due to public access or restricted use, per NRC guidelines, as |

| | locating nuclear plants near these areas can be unacceptable locally. However, like federal lands, protected land can sometimes be used if preferences align with site needs [16]. |
|---|---|
| Population | Nuclear sites must maintain safe distances from populated areas, as outlined in NRC guidelines. Reactors shouldn't be placed near areas where population density exceeds 500 people per square mile or within 4 miles of a town with 25,000 or more residents to minimize radiation exposure risks [17]. |
| Operating Nuclear Facilities | Communities with current nuclear facilities often support new reactor deployment due to job creation, tax revenue, and lower perceived risks. Proximity to existing facilities within 20 miles is measured to assess local support and familiarity with nuclear power [18]. |
| Renewable resources availability | If the area surrounding an existing or retired coal plant has abundant renewable resources like solar or wind, the local community might prefer adopting renewables over new nuclear energy. |

*2.2. Technological Matrix*

| Criteria | Relevance |
|---|---|
| Thermal capacity | A coal to clean energy transition is more favorable when the thermal capacity of the nuclear reactor matches that of the existing or retired coal plant, as it reduces the need for significant technical modifications. |
| Nameplate Capacity | A coal to clean energy transition is more suitable when the nameplate capacity of the coal plant aligns with the nuclear reactor's electric capacity. For advanced reactors, a smaller CPP capacity is preferable. |
| Steam Cycle Temperature and Pressure | The closest match between the steam cycle operating parameters of nuclear and coal power plants will maximize the effective and efficient reuse of coal plant steam cycle components. |
| Fault lines | Nuclear sites must consider faults within 200 miles, as detailed in federal regulations, where fault length dictates the necessary distance. Sites too close to these fault lines are excluded due to seismic safety concerns [19]. |
| Landslide Hazard | NRC guidelines emphasize that nuclear power stations must prevent the loss of safety functions due to ground hazards like earthquakes or landslides. The USGS identifies areas with moderate or high risks for landslides or sinkholes, which need detailed evaluation to ensure safe plant siting [16]. |
| Nuclear Research & Development | Access to nearby nuclear R&D facilities is valuable for technical support when developing new nuclear plants. Institutions like national labs with advanced reactor expertise or universities with research reactors or active nuclear programs can provide vital guidance. Proximity to these R&D hubs is measured by counting facilities within 100 miles of the site [20]. |
| Substations | Electric substations connect power generation sources to the grid, |

| | switch equipment in or out, and adjust voltage levels. The closest substation to the coal site is more suited for coal to clean energy transitions. |
|---|---|
| Generator Retirement | Retiring coal power plants create opportunities for nuclear energy to step in and supply needed electricity while potentially reusing existing infrastructure. This process involves identifying coal plants closing from 2019 to 2050 [21]. For coal to clean energy transitions analysis, the distance from a closing plant within a 1-mile radius of the proposed site can be measured for potential reuse. |
| Transportation | Advanced reactors, often built in factories and transported to their sites, need reliable heavy-haul transport nearby. Thus, transportation means are a crucial factor in the coal-to-nuclear transition. |

### 2.3. *Economic Matrix*

| Criteria | Relevance |
|---|---|
| Energy Price | State electricity prices show where new technology might be cheapest. A higher electricity price allows nuclear and renewable energy to be introduced into the region, making it more promising for the coal to clean energy transition. |
| Net Electricity Imports | States that consume more electricity than they produce are likely to prioritize building new power facilities, such as nuclear and renewables. |
| Market Regulation | Market regulation impacts new reactor siting by influencing costs, financing, and partnerships. Regulated markets let utilities control power generation and delivery, while deregulated markets prevent them from owning both generation and transmission. |
| Construction Labor Rate | Local labor rates impact the construction and maintenance costs of nuclear facilities. |
| Slope | Steep slopes increase the cost of preparing nuclear sites for construction, so large reactor sites should avoid slopes above 12%, per EPRI guidance. SMRs and advanced technologies can handle slopes up to 18% due to their smaller footprints, though the threshold varies based on the technology and economic factors. Sites exceeding these slopes are flagged for further evaluation [4]. |

### 2.4. *Environmental Matrix*

| Criteria | Relevance |
|---|---|
| Safe Shutdown Earthquake | EPRI guidelines advise limiting large LWRs to a safe shutdown earthquake peak ground acceleration of under 0.3 g. In comparison, advanced reactors like SMRs can tolerate up to 0.5 g due to their enhanced seismic resilience. Design features like smaller structures and underground installations improve earthquake resistance [4]. |
| 100 Year Flood | Executive Order 11988 requires government agencies to seek alternatives to development in flood-prone areas. Thus, land within a 100-year floodplain is at risk of adverse effects and should be avoided for new nuclear deployment [22]. |
| Hazardous Facilities | Nearby industrial facilities like chemical plants or oil wells could pose safety risks to nuclear power plants by causing explosions, toxic releases, or fires. To assess their impact on coal to clean energy transitions, it is essential to identify these hazards within five miles of a potential nuclear site [16]. |
| Streamflow | This criterion highlights land areas over 20 miles from freshwater sources suitable for closed-cycle cooling, based on selected makeup water needs. This criterion is unnecessary if once-through or ocean cooling is used, or an advanced reactor technology requires less water. SMRs typically need over 50,000 GPM of cooling water. |
| Open Waters and Wetlands | Open waters, like lakes and rivers, are often protected for drinking, recreation, or navigation and shouldn't be filled to support reactor sites. Wetlands, vital for flood control and biodiversity, are similarly safeguarded under various laws, including the Clean Water Act, which limits intake structures in cooling systems. Federal agencies must avoid developing in wetlands unless no other options are viable, as mandated by Executive Order 11990 [23]. |

## 3. Methodology for evaluating benefits from the long-term planning perspectives

This section aims to assess the techno-economic-environmental feasibility and advantages of retrofitting a coal site for long-term planning. Metrics developed here will compare three energy mix strategies: Greenfield, C2N, and C2IES. To facilitate the comparative analysis for long-term planning, the energy systems will be optimized and evaluated using specific System Performance Indicators (SPIs). The following subsections describe the objective functions, constraints, decision variables, optimization methods, and proposed SPIs for the coal to clean energy transition analysis.

### 3.1. Objective functions

An objective function can either be single-objective or multi-objective. A single-objective function focuses on a single parameter, while a multi-objective function addresses multiple parameters. In a multi-objective function, technical, economic, and environmental parameters can be evaluated simultaneously. The weighted values for each parameter should be selected according to the specific energy infrastructure scenario. Equ. (1) and (2) illustrate the generic forms of single-objective and multi-objective functions for coal to clean energy analysis, respectively.

$$\min F_{sin,obj} = \min\{P(x_1, x_2, \ldots \ldots, x_n)\} \qquad (1)$$
$$\min F_{mul,obj} = \min\{\alpha_1 P_1(x_1, x_2, \ldots \ldots, x_n) + \cdots + \alpha_i P_i(x_1, x_2, \ldots \ldots, x_n)\} \qquad (2)$$

Where, $P$, $x$, and $\alpha$ are the system parameters (techno/economic/environmental), decision variable, and weighted function, respectively.

The following sub-sections detail a list of parameters that can be utilized in the objective function formulation for coal to clean energy transitions.

### 3.1.1. Net Present Cost (NPC)

The Net Present Cost (NPC) is a financial metric used to assess the total cost of an energy system over its lifetime, accounting for both capital investments and operational expenses. It represents the present value of all costs associated with the system, discounted to their equivalent value at the present time. NPC provides insight into the economic viability of the hybrid energy system by quantifying the total financial burden and allowing for comparison with alternative energy solutions.

$$NPC_y = C^y_{cap} + C^y_{OM(f)} + C^y_{OM(v)} + C^y_{rep} + C^y_{fuel} - C^y_{salv} - C^y_{inc}, \qquad \forall y \in \Psi \qquad (3)$$

Where, $NPC_y$, $C^y_{cap}$, $C^y_{OM(f)}$, $C^y_{OM(v)}$, $C^y_{rep}$, $C^y_{fuel}$, $C^y_{salv}$, and $C^y_{PTC}$ are the NPC of the $y$-th component, present worth of capital cost of, present worth of fixed O&M cost, present worth of variable O&M cost, present worth of replacement cost, present worth of fuel cost, present worth of salvage value, and present worth of incentives (e.g., production tax credit) of $y$-th component, respectively.

$$C^y_{cap} = N_y \times C^y_{cl\,(u)} \qquad (4)$$

$$C^y_{OM(f)} = N_y \times C^y_{OM(u)} \times \frac{[(1+d_a)^T] - 1}{d_a(1+d_a)^T} \qquad \forall y \in \Psi \qquad (5)$$

$$C^y_{OM(v)} = N_y \times C^y_{OM(u)} \times \frac{[(1+d_a)^T] - 1}{d_a(1+d_a)^T} \qquad \forall y \in \Psi \qquad (6)$$

$$C^y_{rep} = N_y \times \sum_{W=1}^{W_{nr}} \left[ C^y_{rep\,(u)} \times \frac{1}{(1+d_a)^{(W \times LS_y)}} \right] \qquad (7)$$

$$W_{nr} = \left\lceil \frac{T}{LS_y} \right\rceil - 1 \qquad \{W_{nr} \epsilon Z^+ | W_{nr} \geq 0\} \qquad (8)$$

$$C^y_{fuel} = N_y \times C^y_{fuel(u)} \times \frac{[(1+d_a)^T] - 1}{d_a(1+d_a)^T} \qquad \forall y \in \Psi \qquad (9)$$

$$C^y_{salv} = N_y \times C^y_{rep\,(u)} \times \left[ 1 - \left( \frac{T - LS_y \times \left\lceil \frac{T}{LS_y} \right\rceil}{LS_y} \right) \right] \times \frac{1}{(1+d_a)^T} \qquad \forall y \in \Psi \qquad (10)$$

$$C^y_{inc} = R^y_{inc} \times \frac{[(1+d_a)^T] - 1}{d_a(1+d_a)^T} \qquad \forall y \in \Psi \qquad (11)$$

Where, $N_y$, $Q^y_{cl\,(u)}$, $Q^y_{OM(u)}$, $Q^y_{rep\,(u)}$, $LS_y$, $Q^y_{fuel(u)}$, $R^y_{inc}$, $d_a$, $i$, and $T$ are number of $y$-th component, unit capital cost, unit O&M cost, unit replacement cost, equipment lifespan, unit fuel cost, incentive rate, actual discount rate, inflation rate, and project lifetime, respectively.

For nuclear rector, decommissioning cost will be added to NPC. The decommissioning cost can be expressed as follows.

$$C_{dc}^{Nuc} = Q_{dc(u)}^{Nuc} \times \frac{[(1+d_a)^T]-1}{d_a(1+d_a)^T} \tag{12}$$

Where, $Q_{dc(u)}^{Nuc}$ is the unit decommissioning cost of nuclear reactor.

Additionally, depending on the selected site, transmission lines may need upgrading. This cost is essential for greenfield projects because no existing transmission infrastructure is available. However, in C2N and C2IES scenarios, the requirement for transmission line upgrades should be assessed. Equ. (13) provides an estimation of the transmission line upgrade cost. This equation simplifies the estimation by not accounting for detailed factors influencing transmission upgrade costs, such as transmission distance, required voltage level, terrain type, and other considerations.

$$C_{grid} = a\arctan(N_{grid}+b) + c + d(N_{grid}-N_{grid,min}) \tag{13}$$

Where, $a$, $b$, $c$, and $d$ are the coefficient expressed in \$/MW, MW, \$, and \$/MW, respectively. $N_{grid}$ and $N_{grid,min}$ denote grid interconnection capacity and minimum grid transmission capacity.

### 3.1.2. Annualized Cost of System (ACS)

The Annualized Cost of System (ACS) is a financial metric that represents the yearly equivalent cost of the system, including capital costs, operating expenses, and any financing costs, spread over its lifetime. It accounts for the total investment required to build and operate the hybrid energy system on an annual basis, providing a comprehensive measure of its economic feasibility.

$$CRF = \frac{d_a(1+d_a)^T}{(1+d_a)^T - 1} \tag{14}$$

$$TAC = CRF * NPC \tag{15}$$

Where $CRF$ is the capital recovery factor.

### 3.1.3. Levelized Cost of Energy (LCOE)

The Levelized Cost of Energy (LCOE) is one of the key metrics used to evaluate the cost-effectiveness of electricity generation technologies in energy systems. It represents the average cost of producing electricity over a system's lifetime, considering all capital costs, operating expenses, incentives, and the amount of electricity generated. For an integrated energy system combining multiple sources such as nuclear, renewable, and storage technologies, calculating the LCOE involves summing the present value of all costs and dividing by the total expected energy output. This metric enables the comparison of different energy technologies and informs decision-making when designing efficient and economically viable hybrid energy systems. The LCOE can be expressed as follows.

$$LCOE = \frac{TAC}{Total\ electricity\ produced\ over\ the\ year} \quad (16)$$

### 3.1.4. Loss of Power Supply Probability (LPSP)

The Loss of Power Supply Probability (LPSP) quantifies the probability of insufficient power supply to meet demand during specified periods. It provides insight into the system's reliability and resilience against supply shortages. LPSP can be used either in multi-objective functions or as a constraint. The mathematical expression for LPSP can be given by as follows [25].

$$LPSP = \frac{\sum_{t=1}^{T} |P_{demand}(t) - P_{generation}(t)|}{\sum_{t=1}^{T} P_{demand}(t)} \quad if\ P_{generation}(t) < P_{demand}(t) \quad (17)$$

$$LPSP = \frac{\sum_{t=1}^{T} sign(t)}{T} \quad sign(t) = \begin{cases} 1, & P_{generation}(t) < P_{demand}(t) \\ 0, & P_{generation}(t) \geq P_{demand}(t) \end{cases} \quad (18)$$

Where, $P_{demand}$, $P_{generation}$, and $T$ are the total demand, total generation, and defined timespan, respectively.

### 3.1.5. Total Life Cycle Cost (TLCC)

The Total Life Cycle Cost (TLCC) encompasses all expenses incurred from the system's conception to its decommissioning, including construction, operation, maintenance, and disposal costs. It provides a comprehensive assessment of the financial implications of deploying and operating the energy system over its entire lifespan, aiding decision-makers in evaluating its economic feasibility and long-term sustainability.

## 3.2. Constraints

Constraints encompass technical, economic, and environmental parameters, and in some cases, specific constraints may also be incorporated into the objective function. Below is a list of parameters used in optimizing the coal to clean energy transition analysis.

### 3.2.1. Real estate Constraint (RC)

Since existing or retired coal plants will be used to deploy nuclear reactors and renewable generators, land required for new generators (e.g., nuclear reactors and renewable generators) must be less or equal to the selected coal plant. For greenfield case, this constraint does not exist. For C2N and C2IES, the mathematical expression of this constraint can be given by equ. (19) and equ. (20), respectively.

$$RC_{coal} \geq RC_{nuclear} \tag{19}$$
$$RC_{coal} \geq RC_{IES} \tag{20}$$

Where, $RC_{coal}$, $RC_{nuclear}$, and $RC_{IES}$ are the total area of the coal site, land required for deploying nuclear reactors, and land required for deploying entire IES infrastructure, respectively.

### 3.2.2. Fraction of Load Fulfillment (FLF)

The Fraction of Load Fulfillment (FLF) is a measure used to assess the adequacy of power supply in meeting the electricity demand. Mathematically, FLF can be expressed as:

$$FLF = \frac{\text{Total energy supplied}}{\text{Total energy demanded}} \times 100\% \tag{21}$$

Where, "Total energy supplied" denotes the actual amount of electricity delivered to consumers over a specified period and "Total energy demanded" represents the total electricity demand from consumers over the same period. FLF quantifies the proportion of the electricity demand that is met by the available power supply. A higher FLF indicates a greater fulfillment of the electricity demand, while a lower FLF suggests an inadequate supply relative to demand. FLF will be 1 if demand exactly matches with the supply/generation.

### 3.2.3. Excess Energy Fraction (EEF)

The Excess Energy Fraction (EEF) is a measure used to quantify the surplus energy generated by a power system compared to the electricity demand. EEF can vary from 10-59% [26], [27]. In some literature, research recommends keeping EEF less than 10% [28]. However, it depends on the specific case and the system's requirements. Mathematically, EEF can be expressed as:

$$EEF = \frac{\sum_{t=1}^{T}|P_{generation}(t) - P_{demand}(t)|}{\sum_{t=1}^{T}P_{demand}(t)} \qquad if\ P_{demand}(t) < P_{generation}(t) \tag{22}$$

### 3.2.4. Energy Not Supplied (ENS)

Energy Not Supplied (ENS) is a metric (expressed in kWh or MWh) used to quantify the amount of electricity demand that remains unmet due to insufficient power supply. Mathematically, ENS can be expressed as [29]:

$$ENS = \sum_{t=1}^{T} E_{unmet}(t) \tag{23}$$

Where, $E_{unmet}$ is the unserved energy at t-th hour in timespan T. It provides insight into the extent to which the power system fails to meet the demand, indicating potential energy shortages or reliability issues. It can also be used as SPI.

### 3.2.5. Renewable Fraction (RF)

The Renewable Fraction (RF) quantifies the proportion of electricity generated from renewable energy sources relative to the total electricity generated. RF can be represented mathematically as follows:

$$RF = \frac{\sum_{t=1}^{T} P_{total}(t) - \sum_{t=1}^{T} P_{nuclaer}(t)}{\sum_{t=1}^{T} P_{total}(t)} \tag{24}$$

Where, $P_{total}(t)$ and $P_{nuclaer}(t)$ represent total power generation from all resources within timespan T and power generation form nuclear reactor within timespan T, respectively. It can be used for the C2IES case since greenfield and C2N do not include any renewable energy resources. If someone wants to limit the renewable contributions at a specific level due to uncertainty, it might be an essential constraint for that case. Since a high level of renewable penetration may cause the system to be unreliable, it is a critical constraint for the C2IES case.

### 3.2.6. Energy balance constraint

The Energy Balance Constraint ensures that the total electricity generated by the power system is equal to the total electricity consumed or demanded. It can be illustrated by the following equation.

$$\sum_{t=1}^{T} \sum_{i=1}^{n} P_{gen,i}(t) = \sum_{t=1}^{T} \sum_{j=1}^{m} P_{con,j}(t) \tag{25}$$

Where, $P_{gen,i}(t)$ and $P_{con,j}(t)$ denotes power generation by i-th generation source at t-th hour and power consumption by j-th consumer/load source at t-th hour, respectively.

### 3.2.7. Equipment operational constraint

Equipment operational constraints are limitations on the operation of specific components or units within a power system. These constraints ensure that equipment operates within safe and efficient operational boundaries. Equipment operational constraints can be expressed using inequalities or equations that specify acceptable ranges for parameters such as power output or operating hours. For example, an operational constraint for a generator G could be represented as:

$$P_{min,G} \leq P_G \leq P_{max,G} \tag{26}$$

Where, $P_{min,G}$, $P_G$, and $P_{max,G}$ present minimum output power of generator G, power output of generator G, and maximum output power of generator G, respectively.

### 3.3. Decision variables

Decision variables vary based on system configuration. A list of potential decision variables for the energy system model in coal to clean energy transitions is provided below. Not all decision variables need to be included in a single system; for instance, if the proposed system is not grid-connected, grid sale/purchase variables will be excluded from the optimization problem.

| Decision variables | Notation | Explanation | Unit | Applicability |
|---|---|---|---|---|
| Nuclear reactor ($N_{nuc}, Cap_{nuc}$) | $N_{nuc}^{min} \leq N_{nuc} \leq N_{nuc}^{max}$ | $N_{nuc}^{min}$: Minimum number of reactors $N_{nuc}^{max}$: maximum number of reactors | Quantity | Greenfield, C2N, C2IES |
| | $Cap_{nuc}^{min} \leq Cap_{nuc} \leq Cap_{nuc}^{max}$ | $Cap_{nuc}^{min}$: Minimum capacity of nuclear generation $Cap_{nuc}^{max}$: maximum capacity of nuclear generation | $kW_e/MW_e/GW_e$ | |
| PV Panels ($N_{PV}, CAP_{PV}$) | $N_{PV}^{min} \leq N_{PV} \leq N_{PV}^{max}$ | $N_{PV}^{min}$: Minimum number of PV panels $N_{PV}^{max}$: Maximum number of PV panels | Quantity | C2IES |
| | $Cap_{PV}^{min} \leq CAP_{PV} \leq Cap_{PV}^{max}$ | $Cap_{PV}^{min}$: Minimum capacity of solar generation $Cap_{PV}^{max}$: Maximum capacity of solar generation | $kW_e/MW_e$ | |
| WT ($N_{WT}, CAP_{WT}$) | $N_{WT}^{min} \leq N_{WT} \leq N_{WT}^{max}$ | $N_{WT}^{min}$: Minimum number of WT $N_{WT}^{max}$: Maximum number of WT | Quantity | C2IES |
| | $Cap_{WT}^{min} \leq CAP_{WT} \leq Cap_{WT}^{max}$ | $Cap_{WT}^{min}$: Minimum capacity of wind generation $Cap_{WT}^{max}$: Maximum capacity of wind generation | $kW_e/MW_e$ | |
| Battery storage ($N_{BS}, CAP_{BS}$) | $N_{BS}^{min} \leq N_{BS} \leq N_{BS}^{max}$ | $N_{BS}^{min}$: Minimum number of battery bank $N_{BS}^{max}$: Maximum number of battery | Quantity | Greenfield, C2N, C2IES |

| | | bank | | |
| --- | --- | --- | --- | --- |
| | $Cap_{BS}^{min} \leq CAP_{BS} \leq Cap_{BS}^{max}$ | $Cap_{BS}^{min}$: Minimum capacity of battery storage<br>$Cap_{WT}^{max}$: Maximum capacity of battery storage | kWh/MWh | |
| Hydrogen tank ($N_{HT}$) | $N_{HT}^{min} \leq N_{HT} \leq N_{HT}^{max}$ | $N_{HT}^{min}$: Minimum number of hydrogen tank<br>$N_{HT}^{max}$: Maximum number of hydrogen tank | Quantity | Greenfield, C2N, C2IES |
| Electrolyzer ($Cap_{elec}$) | $Cap_{elec}^{min} \leq Cap_{elec} \leq Cap_{elec}^{max}$ | $Cap_{elec}^{min}$: Minimum capacity of electrolyzer<br>$Cap_{elec}^{max}$: Maximum capacity of electrolyzer | kW/MW | Greenfield, C2N, C2IES |
| Fuel cell ($Cap_{FC}$) | $Cap_{FC}^{min} \leq Cap_{FC} \leq Cap_{FC}^{max}$ | $Cap_{FC}^{min}$: Minimum capacity of fuel cell<br>$Cap_{FC}^{max}$: Maximum capacity of fuel cell | kW/MW | Greenfield, C2N, C2IES |
| Thermal energy storage | $Cap_{TES}^{min} \leq CAP_{TES} \leq Cap_{TES}^{max}$ | $Cap_{TES}^{min}$: Minimum capacity of TES<br>$Cap_{TES}^{max}$: Maximum capacity of TES | kWh/MWh/ GWh | Greenfield, C2N, C2IES |
| Grid sale | $Cap_{GS}^{min} \leq CAP_{GS} \leq Cap_{GS}^{max}$ | $Cap_{GS}^{min}$: Minimum power selling capacity of grid<br>$Cap_{GS}^{max}$: Maximum power selling capacity of grid | kW/MW/GW | Greenfield, C2N, C2IES |
| Grid purchase | $Cap_{GP}^{min} \leq CAP_{GP} \leq Cap_{GP}^{max}$ | $Cap_{GP}^{min}$: Minimum power purchasing capacity of grid<br>$Cap_{GP}^{max}$: Maximum power purchasing capacity of grid | kW/MW/GW | Greenfield, C2N, C2IES |

### 3.4. Optimization techniques

Optimization techniques can be categorized into Classical, Artificial intelligence, Probabilistic, and Hybrid methods [30], [31].

3.4.1. Classical optimization techniques

Classical optimization methods refer to traditional optimization techniques used for decades to solve optimization problems. These methods rely on mathematical formulations and algorithms to find the optimal solution to a given problem. Classical optimization methods include Linear Programming (LP), Nonlinear Programming (NLP), Integer Programming (IP), Dynamic Programming (DP), Analytical, Numerical, Graphical Construction, Iterative, and Convexification methods. Classical optimization methods are based on well-established mathematical principles, providing a rigorous framework for solving optimization problems. Many classical optimization algorithms have been optimized for efficiency and can handle large-scale optimization problems effectively. Classical optimization methods typically converge to deterministic solutions, providing transparent and interpretable results. Therefore, classical optimization methods have limited application in solving long-term planning problems concerning coal to clean energy analysis due to a large number of decision variables and constraints.

### 3.4.2. Artificial intelligence techniques [31]

Artificial intelligence and nature-based (AIN) optimization techniques leverage computational intelligence and machine learning algorithms to solve optimization problems. These methods often exhibit greater flexibility and adaptability compared to classical optimization methods. AIN optimization methods include Genetic Algorithms, Cross Entropy, Simulated Annealing, Particle Swarm Optimization, Artificial Bee Colony, Biogeography-bases Optimization, Flower pollination, Social Spider Optimization, Grey Wolf Optimization, Jaya Algorithm, Dragonfly Algorithm, Pity Beetle Algorithm, Coyote Optimization, Deer Hunting Optimization, Forensic-based Investigation Algorithm, manta Ray Foraging, Golden Eagle Optimizer, Tunicate Swarm, and Jellyfish Optimizer. AIN optimization methods can adapt to various problem structures, including nonlinearities, discrete decisions, and uncertainties, making them suitable for complex optimization tasks. Some AIN techniques, such as GA and PSO, are capable of exploring the entire solution space to find global optima, avoiding the issue of getting stuck in local optima. Techniques like Artificial Neural Networks (ANN) and Deep Reinforcement Learning (DRL) can learn from data and experience, enabling them to improve performance over time and handle dynamic or uncertain environments. However, AIN optimization methods often require careful tuning of parameters, and their performance may be sensitive to parameter settings or initialization conditions. Therefore, AIN optimization techniques are extensively used for solving energy system optimization problems, particularly in cases involving complexity, uncertainty, and nonlinearity. It is, thus, highly recommended for long-term planning problems of coal to clean energy transition analysis.

### 3.4.3. Probabilistic method

Probabilistic optimization techniques incorporate probabilistic models or uncertainty considerations into the optimization process. These techniques are beneficial for solving optimization problems in which parameters or constraints are uncertain or subject to variability. Probabilistic optimization techniques include Stochastic Programming, Chance-Constrained Programming, Robust Optimization, Monte Carlo Simulation, and Bayesian Optimization. Probabilistic optimization techniques are used to solve hybrid energy system optimization problems by explicitly considering uncertainties related to renewable energy availability, demand variability, and equipment reliability. By accounting for uncertainties, these techniques help identify robust solutions that perform well under different operating conditions or future scenarios. Probabilistic techniques produce robust solutions against uncertainties, ensuring reliable performance in real-world situations. Probabilistic techniques allow decision-makers to assess and manage risks associated with uncertainties, enabling better-informed decisions. However, analyzing uncertainties and exploring multiple scenarios can increase computational demands, especially for large-scale problems or complex models. Besides, the effectiveness of probabilistic techniques depends on the accuracy of the probabilistic models and assumptions made about uncertainty distributions, which may introduce errors or biases. Therefore, C2IES may require probabilistic optimization techniques to assess the impact of renewables on the system's techno-economic viability.

### 3.4.4. Hybrid methods

The hybrid optimization technique combines various algorithmic approaches to find the best solution for managing and optimizing hybrid energy systems. Hybrid optimization techniques include Heuristic Combination Algorithms, Fuzzy Logic/Iteration + Heuristic, Probability + Conventional Methods, and other newer or less common approaches such as machine learning-based optimizations, neural networks, or agent-based modeling. Hybrid optimization techniques provide consistent results under similar conditions. They are often scalable for large-scale problems, given sufficient computational resources. However, they might not handle the non-linearities and complexities of modern hybrid energy systems effectively. They can be computationally intensive, especially for methods like linear programming on large-scale systems. Therefore, a case-specific decision is required on whether we should utilize hybrid optimization techniques in coal to clean energy transition analysis.

### 3.5. System Performance Indicator (SPI)

A list of SPI is listed below.

| SPI | Definition | Expression |
|---|---|---|
| Deficit Power Probability (DPP) [32] | Probability that the power generation will fall below the load demand at any given time | $DPP = \dfrac{\text{number of time intervals for power deficiency}}{\text{total number of time intervals considered}}$ |
| Loss of Load Expected (LOLE) [33] | Number of hours when load is expected to exceed available generation capacity | $LOLE = \sum_{t=1}^{T} t_{outage}$ |
| Self-consumption Ratio (SCR) of reactor [34] | Quantification of reactor contribution while serving the total system demand | $SCR = \dfrac{\sum_{t=1}^{T} P_{nuclear}(t)}{\sum_{t=1}^{T} P_{load}(t)}$ |
| Load Dissatisfaction | Percentage of energy deficiency of the system | $LDR = \dfrac{\sum_{t=1}^{T} P_{load}(t) - \sum_{t=1}^{T} P_{supply}(t)}{\sum_{t=1}^{T} P_{load}(t)} \times 100\%$ |

| Metric | Description | Formula |
|---|---|---|
| Rate (LDR) [35] | | |
| Grid Dependency (GD) [36] | Quantification of imported grid energy compared to total demand | $GD = \dfrac{Total\ enery\ imported\ from\ grid}{Total\ demand} \times 100\%$ |
| Seasonal Loss of Load Probability Ratio (SLLPR) | Quantification of how the LOLP during a particular season compares to the LOLP averaged over the entire year. | $SLLPR = \dfrac{\text{LOLP during a specific season}}{\text{LOLP averaged over the entire year}}$ |
| Level of Autonomy (LA) [37] | Quantification of the duration when loss of load does not occur | $LA = 1 - \dfrac{No.\ of\ hours\ when\ loss\ of\ load\ occurs}{Total\ number\ of\ operational\ hours}$ |
| Human Development Index (HDI) [38] | Quantification of a country's social and economic development level. $E_{cons\_annual\_per\_capita}$ denotes annual electricity consumption per capita (kWh/yr/person) | $HDI = [0.0978 \times \ln(E_{cons\_annual\_per\_capita})] - 0.0319$ |
| Job Creation (JC) [38] | Quantification of job creation by different system components adopted. $jc$ and $P$ represent number of jobs per MWp of system components (e.g., reactor, renewable generators, and other components) and peak capacity of the system component. | $JC = \sum_{m=1}^{M} jc_m P_{renew,m} + jc_{nuc} P_{nuc} + \sum_{n=1}^{N} jc_n P_{com,n}$ |
| Life Cycle Assessment (LCA) [38] | Quantification of the greenhouse gas emissions. For C2N and C2IES, emissions during CPP demolition and ash removal should be added. | $LCA = \left[\sum_{m=1}^{M} \sum_{t=1}^{T} EM_m P_m(t) N_m \Delta t\right]$ $+\ emisison\ due\ to\ \text{demolition and ash removal}$ |
| Grid Interaction Level (GIL) [39] | Evaluation of how frequent energy system interacts with the grid since frequent | $GIL = \dfrac{\left|E_{grid}^{in}\right| + \left|E_{grid}^{out}\right|}{E_{system}^{load}} \times 100\%$ |

| | interaction may cause instability. $E_{grid}^{in}$ and $E_{grid}^{out}$ refer to the energy purchased from and sell to the grid. $E_{system}^{load}$ is the total system load. | |
| Avoided Carbon Footprint (ACF) | Quantification of the amount of carbon emissions that can be avoided by adopting proposed system (e.g., greenfield, C2N, C2IES) | $ACF = LCA_{coal} - LCA_{proposed\_sys}$ |

## 4. Methodology for evaluating benefits from short-term operational perspectives

This segment of the analysis aims to evaluate the comparative operational (short-term) benefits of three proposed systems (Greenfield, Brownfield/C2N, and C2IES) alongside a hypothetical operational coal plant. To facilitate this comparison, a Unit Commitment Economic Dispatch (UCED) problem can be formulated and solved for all four cases to assess their techno-economic advantages. The following subsection outlines the objective function, constraints, and decision variables. The optimization methods that can be used to solve the problem have already been explained in section 3.4. It is not necessary to include every constraint and decision variable in a single optimization problem, as they can be adapted based on design requirements. Additionally, this study should consider a test power system with generators, loads, and transmission lines.

### 4.1. Objective function

The objective of the UCED problem here is to minimize the operating cost of the energy system while adhering to all constraints. Operating costs in the U.S. electricity market usually encompass energy and reserve capacity costs. The objective function can be formulated as follows.

$$minimize \: \{f^g(P_g, w_g) + G^g(z_g)\} + f^r(R_g) + f^{ES}(R_{ES}) \tag{27}$$
$$f^g(P_g, w_g) = k_g^{fuel}(a_g P_g + b_g w_g) + k_g^{vom} P_g \tag{28}$$
$$G^g(z_g) = S_g z_g \tag{29}$$

Where, $f^g(P_g, w_g)$, $f^r(R_g)$, $G^g(z_g)$ and $f^{ES}(R_{ES})$ refer to system operating cost due to energy, reserve, generator start-up cost, and energy storage, respectively. $P_g, w_g, k_g^{fuel}, k_g^{vom}, z_g$, and $S_g$ denote power generation of each generator, unit commitment status [0, 1], fuel price of generator, variable O&M cost of generator, start-up status [0, 1], and start-up cost of generator, respectively. $a_g$ and $b_g$ are the linear cost curve coefficient of generators. For coal to clean energy transition study, the generators are coal, nuclear reactor, and renewable generators (e.g., WT and solar PV panels). The energy storage may include battery storage, fuel cell, and flywheel. The inclusion of system component depends on system modelers. A liner cost function for generators has been assumed here. It can be replaced by available quadratic cost function.

### 4.2. Constraints

One of the crucial constraints in this study is the formulation of power flow equations. Power flow equations are typically nonlinear, making the problem highly complex and time-consuming, especially for larger power systems. Hence, linearizing these nonlinear equations (known as DC power flow equations) is often advisable to simplify the optimization process. DC power flow equations yield reasonable results compared to their nonlinear counterparts and are widely used in power system operations. As a result, this study adopts DC power flow equations. A list of potential constraints is provided below.

#### 4.2.1. Power flow constraint

The DC power flow equations can be expressed as follows:

$$P_f = B_f \Theta \quad (30)$$
$$P_{bus} = B_{bus} \Theta \quad (31)$$

Equ. (30) is the DC power flow equation while equ. (31) determines the net power injection from voltage angles. $P_f$, $\Theta$, and $P_{bus}$ denote power flow through each transmission line, voltage angle at each bus, and power injection into each bus, respectively. $B_f$ and $B_{bus}$ are the susceptance matrices of the transmission lines.

#### 4.2.2. Reserve requirement constraint

The reserve requirement of the system can be written as follows.

$$\mathbf{1}^T R_g + \mathbf{1}^T R_{ES} \geq r_{req} \quad (32)$$

Where, $\mathbf{1}^T$, $R_g$, $R_{ES}$, and $r_{req}$ denote vector of ones of appropriate dimension, reserve capacity of each generator, reserve capacity of each energy storage, and required reserve capacity by the energy system, respectively.

#### 4.2.3. Generator, storage, transmission line limit constraint

The generators and energy storage operational limits can be presented as follows.

$$P_g - R_g \geqslant P_g^{min} \quad (33)$$
$$P_g + R_g \leqslant P_g^{max} \quad (34)$$
$$R_{ES}^{min} \leqslant R_{ES} \leqslant R_{ES}^{max} \quad (35)$$
$$P_f^{min} \leqslant P_f \leqslant P_f^{max} \quad (36)$$

Here, $\geqslant$ and $\leqslant$ refer to elementwise inequalities. $P_g^{min}$, $P_g^{max}$, $R_{ES}^{min}$, $R_{ES}^{max}$, $P_f^{min}$, $P_f$, and $P_f^{min}$ denote minimum generation level of each generator, maximum generation level of each generator, minimum reserve capacity of each storage, maximum reserve capacity of each storage, minimum power flow of each line, power flow of each line, and maximum power flow of each line, respectively. The above equations are formulated by considering symmetrical up and down reserve capacity.

### 4.2.4. Minimum up-time constraint

Operators may choose that a few types of generators (e.g., nuclear reactors) should continue to supply for a minimum definite time once they come online. This type of constraint can be formulated as follows.

$$\sum_{k=t-\alpha+1}^{t} z_g^k \leq w_g^k \tag{37}$$

Where $\alpha$ is the minimum up-time.

### 4.2.5. Ramp-up and ramp-down constraint

This constraint defines the ramp rate of each generator. While ramp rate is less crucial for renewable generators, it is particularly important for flexible nuclear reactors and coal-fired generators. The ramp-up and ramp-down constraints for any generator can be expressed as follows.

$$P_g(t) - P_g(t-1) \leq Ramp_{up} \tag{38}$$
$$P_g(t-1) - P_g(t) \leq Ramp_{down} \tag{39}$$

Where, $Ramp_{up}$ and $Ramp_{down}$ are the ramp-up and ramp-down rate of the generator.

### 4.2.6. Reactor minimum stable period

For flexible nuclear reactors, it's advisable to remain in a stable period after ramp-down for a specific time (e.g., a few hours) to reduce xenon concentration. This physical constraint should be applied based on the reactor type. However, it can be disregarded for Fast Neutron Reactors (FNRs) since they are minimally affected by xenon poisoning [40].

### *4.3. Decision variables*

The decision variables can be represented as follows. It's important to note that the number of decision variables may vary based on the problem formulation and system architecture. The decision variable obtained for each (total four cases) will be inserted into the objective function to compare the cost-effectiveness of each option.

$$X = [P_g, \Theta, R_g, R_{ES}, w_g, z_g] \tag{40}$$

## 5. Uncertainty and sensitivity analysis

The uncertainty analysis is particularly important for long-term planning and short-term operational problem (Section 3 and 4, respectively), while sensitivity analysis is important for all three sections (Section 2, 3, and 4).

### *5.1. Uncertainty analysis*

In long-term planning, annual load and renewable resource data (such as solar irradiance and wind speed) are typically gathered and incorporated into the optimization problem. These data are often assumed to be perfectly predicted over the long term (20, 40, or 60 years), which is not conservative. Therefore, it is essential to model load profiles and resource data probabilistically to capture uncertainties in each data type [41]. In the probabilistic approach, resource and load profiles are selected randomly based on probability values and then used in optimization. Optimization techniques that account for uncertainty include robust optimization, stochastic programming, fuzzy programming, and interval methods [42].

### 5.1.1. Modeling of renewable energy resources

A Probability Distribution Function (PDF) must be formulated to introduce uncertainty through a probabilistic approach. PDFs are generally derived from collected solar irradiance and wind speed data. While a beta distribution is often employed for solar irradiance, other distributions such as Weibull, lognormal, inverse-Gaussian, and gamma distribution can also be used to model its PDF [43]. Similarly, wind speed can be modeled using Weibull or Rayleigh distribution functions [44].

### 5.1.2. Modeling of demand profile

Modeling demand or load profiles in power systems commonly involves various PDFs to capture the variability and uncertainty inherent in electricity consumption accurately. The PDFs include Normal distribution, Lognormal distribution, Gamma distribution, Weibull distribution, and Poisson distribution [45], [46].

## 5.2. Sensitivity analysis

Sensitivity analysis examines how each system parameter affects the research findings. In this analysis, a single parameter is adjusted while keeping the others constant to measure its impact on the final results. Since socio-techno-economic-environmental analyses of coal to clean energy transitions involve numerous economic and technical parameters, sensitivity analysis is essential for understanding how system performance changes with varying parameters. However, there are no universally recommended parameters listed in Sections 2, 3, and 4 for sensitivity analysis, as it depends entirely on the specific system models.

## 6. Conclusions and future works

This paper presents a comprehensive methodology for evaluating coal to clean energy transitions using the Complex Nexus of STEEP metrics. Our methodological approach is divided into three parts: optimal site selection, long-term planning assessment, and short-term operational benefits evaluation. The optimal site selection methodology includes societal, technological, economic, and environmental metrics, while the long-term planning methodology focuses on technical, financial, and environmental criteria. Since energy system operations prioritize minimizing total operating costs while maintaining technical constraints, short-term operational benefits are typically assessed based on technical and economic metrics. The study also includes uncertainty and sensitivity analyses, which are crucial for coal to clean energy analysis. It is ultimately up to energy system modelers to decide where and how to incorporate uncertainty and sensitivity assessments.

Future research will implement all three types of metrics separately into computer simulations. Developing simulation tools to validate and test these frameworks across diverse scenarios will enhance their applicability in real-world transitions. Future research should also explore refining these methodologies further to incorporate emerging nuclear technologies and region-specific regulatory constraints. Additionally, integrating more granular data on socio-economic factors and public acceptance would help in crafting nuanced strategies. Research should also address uncertainty modeling and sensitivity analysis to ensure that these transitions are resilient to future shifts in market dynamics and policy landscapes. Ultimately, continued collaboration between researchers, industry, and policymakers will be vital for successfully implementing these strategic methodologies to achieve a sustainable energy future.

## Acknowledgement

This work is sponsored by the Department of Energy Office of Nuclear Energy under project number (DE-NE0009382), which is funded through the Nuclear Energy University Program (NEUP).

## CRediT author statement

**Muhammad R. Abdussami**: Conceptualization, Methodology, Software, Validation, Formal analysis, Investigation, Resources, Data Curation, Writing - Original Draft, Writing - Review & Editing, Visualization. **Aditi Verma**: Conceptualization, Validation, Writing - Review & Editing, Visualization, Investigation, Supervision, Project administration.